# Quantum theory from Hamilton's Principle with imperfect information

John Hegseth*
Department of Physics
University of New Orleans
New Orleans, LA 70148

Many quantization schemes rely on analogs of classical mechanics where the connections with classical mechanics are indirect. In this work I propose a new and direct connection between classical mechanics and quantum mechanics where the quantum mechanical propagator is derived from a variational principle. This principle allows a physical system to have imperfect information, i.e., there is incomplete knowledge of the physical state, and many paths are allowed.

Physical laws expressed as variational principles are appealing for their simplicity and generality [1], [2], [3]. In addition, they often provide information about a system's stability based on the type of extrema (e.g., minima, maxima, etc.) [4]. Such a global character often leads us to anthropomorphic interpretations where we assign a goal or motive to the behavior. In classical physics the type of extrema depends on the details of a specific system's Hamiltonian [2], [5]. In classical physics the action $S[p(t),q(t),t] = \int (p\dot{q} - H(p,q))dt$ is extremized. Another action $R[p,q,t] = \int (-q\dot{p} - H(p,q))dt$ may also be used to derive Hamilton's equations or a Hamilton-Jacobi equation [5]. Both $R$ and $S$ require perfectly known beginning ($t_i$) and ending ($t_f$) conditions for extremization: $\delta p(t_i)=0$, $\delta p(t_f)=0$ and $\delta q(t_i)=0$, $\delta q(t_f)=0$ respectively. In this paper I propose another variational principle that describes the motion when perfect information about $p$ and $q$ is not available. This is done by generalizing both $S$ and $R$ to include all possible paths and introducing two distributions functionals over the possible $p(t)$ paths $\alpha[p(t)]$ and the possible $q(t)$ paths $\beta[q(t)]$. I will first discuss these mixed path distributions and how they are normalized for fermions and bosons. I will next show how Hamilton's principle is generalized and how the probability amplitude naturally follows.

Let experimenters $A$ and $B$ observe a system of particles. $A$ is only capable of sampling the momentum path $p(t)$ and $B$ is only able to measure points on the position path $q(t)$. Between sampling time interval $\Delta t$, $A$ may observe an initial and final $p$ and extremizes $R$. During $\Delta t$ $B$ may observe an initial and final $q$ and extremizes $S$. If these initial and final points are not known, then there are many possible paths. Figure 1 shows an example, during two time intervals, where $q_i$ is known but there are three possible $p_i$ values that $A$ could observe. If either experimenter simultaneously observed both $p$ and $q$, then the behavior of the system could be determined. In addition, there are points in the system at a given initial time $t_i$ and a given final time $t_f$ that are known by $A$ or $B$ (e.g., they are measured). Points that could be measured but are not measured are not known by $A$ or $B$.

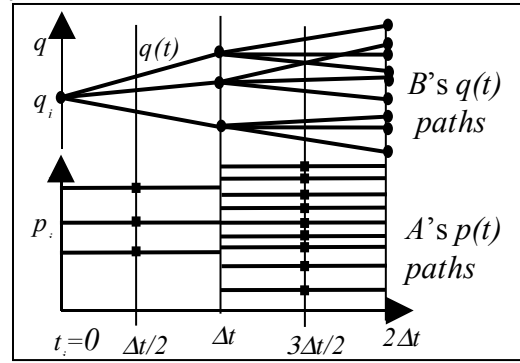

Figure 1. Because there are three possible initial $p$ values, the number of possible paths increases geometrically with time (paths fan out). Each $q(t)$ path that $B$ could observe implies a corresponding $p(t)$ path through Hamilton's equations (and vice versa for $A$) To avoid simultaneous knowledge of $p$ and $q$ $A$'s and $B$'s sampling time are offset by $\Delta t/2$.

Clearly there is a very significant increase in the number of possible paths in time as illustrated in Figure 1. If either $A$ or $B$ fixed an endpoint at a later time $t_f$, e.g., the fixed endpoint $q(t_f)$, the number of possibilities would be reduced.

I next define the distributions for these possible paths: $\alpha[p(t)]$ for $A$ and $\beta[q(t)]$ for $B$ between fixed beginning and ending points. It is important to note that for each possible complete path $p(t)$ (or $q(t)$) there is a corresponding $q(t)$ (or $p(t)$). For each possible $q(t)$ and $\dot{q}(t)$ in $B$'s set of possible paths, he can infer a possible $p(t)$ using Hamilton's equation $\dot{q} = H_p$. Similarly, for each possible $p(t)$ and $\dot{p}(t)$ in $A$'s set, he can infer a $q(t)$ path using $\dot{p} = -H_q$. I note that this is a relation between *possible* paths and that there is a one-to-one relation between possible $p(t)$'s and $q(t)$'s. Because both $A$ *and* $B$ could measure or infer $q(t)$, the probability for a given $q(t)$ is $\beta[q]\beta[q] = \beta^2[q]$. Similarly, the probability for a given $p(t)$, that could be measured by $A$ *and $B$*, is $\alpha[p]\alpha[p] = \alpha^2[p]$. The probability $P_p$ for a particular path between specified points at $t_i$ and $t_f$, e.g., $(q_i, t_i)$ and $(q_f, t_f)$, is the probability that $A$ *and $B$* could observe the path in terms of $q(t)$ *or* that $A$ *and $B$* could observe the path in terms of $p(t)$, i.e., $P_p = \beta[q]\beta[q] + \alpha[p]\alpha[p]$. I note that $\beta[q]$ may be negative and an interpretation of $B$'s distribution $\beta[q]$ (or $A$'s distribution $\alpha[p]$) as probabilities is unnecessarily restrictive. I call the distributions $-1 \leq \beta \leq 1$ and $-1 \leq \alpha \leq 1$ mixed paths. $\alpha^2$ (or $\beta^2$) is the probability that $A$ and $B$ could observe a given $p(t)$ (or $q(t)$).

I now consider many possible paths for bosons. A similar argument can also be made for fermions. Let $B$ hypothesize a given path, say $q_1$. $A$ could measure any of the possible paths $q_1, q_2, ..., q_n$. The probability that $B$ could observe $q_1$ while $A$ observes any of the paths in terms of $q$ is $\beta[q_1]\Sigma_j\beta[q_j]$ for $j=1,2,...,n$. The probability for $A$ and $B$ to observe any of the paths in terms of $q$ is $\Sigma_k\beta[q_k]\Sigma_j\beta[q_j] = \{\Sigma_j\beta[q_j]\}^2$. Similarly, the probability for $A$ and $B$ to observe any of the paths in terms of $p$ is $\{\Sigma_j\alpha[p_j]\}^2$. The total probability $P_T$ to observe any of the possible paths between given beginning and ending spatial points is $P_T(q_i,q_f) = \{\Sigma_j\alpha[p_j]\}^2 + \{\Sigma_j\beta[q_j]\}^2$. In terms of the relative probability with respect to $P_T$, I get the normalization $\{\Sigma_j\alpha[p_j]\}^2 + \{\Sigma_j\beta[q_j]\}^2 = 1$. I can now allow a fan out between paths at any instant and any position in an interval $T = n\Delta t$ by letting $\Delta t \to 0$ and $n \to \infty$ to define a path integral $\Sigma_j\alpha[p_j] \to \int Dp\alpha[p(t)]$ and $\Sigma_j\beta[q_j] \to \int Dq\beta[q(t)]$, where $Dq$ and $Dp$ are the measures of the integral. I can find the probability $P_T$ for a particle to start at $q_i$ and end at $q_f$ by summing over all possible paths or performing the path integrals $\int Dq(t)$ and $\int Dp(t)$ to get $P_T(q_i,q_f) = \{\int Dp\alpha[p(t)]\}^2 + \{\int Dq\beta[q(t)]\}^2$.

$A$ and $B$ may measure their respective *rotating* variables if the particle has integer spin (e.g., $B$ is capable of measuring the angle variable using polarizing filters). If a particle has an internal ½ integer spin degree of freedom, however, as discussed in reference [5], then $A$ and $B$ are only able to actually measure $p(t)$ (e.g., measure $p = \pm constant$ through an electron's magnetic moment). The distributions $\alpha[p(t)]$ for $A$ and $\beta[q(t)]$ for $B$ are observable between $(p_i, p_f)$ and not between the angle variables $(q_i, q_f)$ that can not be fixed in advance as they are impossible to observe. Because $p = \pm const.$, $\dot{p}$ is undefined and $A$ cannot invert $\dot{p} = -H_q$ to find a corresponding $q(t)$. A specific $q(t)$ cannot be inferred from a specific $p(t)$. A given $q(t)$ can still be hypothesized and $\beta[q(t)]$ still exists as it is defined over all possible $q(t)$. Because $A$ could not measure or infer $q(t)$, if he could measure $p(t)$, the probability for a given $q(t)$ that could be measured by $A$ and $B$ is $\beta^2[q] = 0$. In other words, it is not possible for $A$ and $B$ to measure a given $q(t)$. Because a possible $q(t)$ would have cusps corresponding to $p(t)$ discontinuities, $B$ can't invert $\dot{q} = H_p$. Because $B$ could not infer $p(t)$, the probability for a given $p(t)$, measured by $A$ *and $B$*, is $\alpha[p]\alpha[p] = \alpha^2[p] = 0$. Because the classical $p(t)$ and $q(t)$ paths exist in principle, the corresponding distributions $\alpha[p]$ and $\beta[q]$ must not be zero. They must be anti-commuting Grassman numbers: $\alpha[p]\beta[q] = -\beta[q]\alpha[p]$. The measurement of $p$ and $q$ is no longer mutually exclusive and the probability $P_p$ for a particular path between $(p_i, t_i)$ and $(p_f, t_f)$, is no longer $\beta^2 + \alpha^2$, i.e., $\beta[q]\beta[q] + \alpha[p]\alpha[p] = 0$. In fact, this is a statement that such an observation process is impossible. It is possible, however, for $A$ to represent a path in terms of $p(t)$ and $B$ to represent a path in terms of $q(t)$, i.e., $\alpha[p]\beta[q] \neq 0$. It is impossible for $A$ and $B$ to represent the path in terms of either $p(t)$ and $q(t)$ or $q(t)$ and $p(t)$: $\alpha[p]\beta[q] + \beta[q]\alpha[p] = 0$. We may then interpret the positive probability, such as $\alpha[p]\beta[q]$, as the probability of a possible outcome as usual. The negative quantity $\beta[q]\alpha[p]$, is not just impossible but anti-possible, i.e., it is an independent alternative that annihilates a possible outcome to produce a normalization statement of absolute certainty of impossibility. These events together are impossible because of the broken classical connection between $p$ and $q$ that results from the absence of possible information about $q$. By generalizing the mixed path to include Grassman numbers, the anti-commuting behavior of fermions is described. $\alpha\beta$ is the probability for a given path and $P_T(p_i, p_f) = \int Dp \int Dq \, \alpha[p(t)]\beta[q(t)]$.

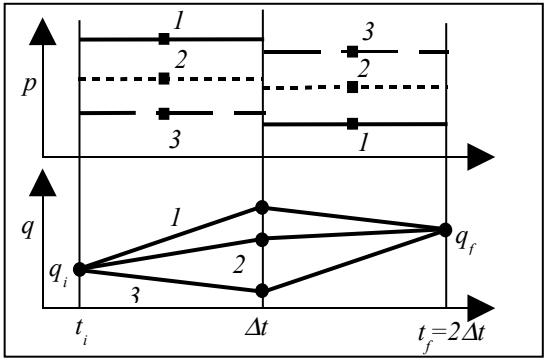

Figure 2. Five spatial-temporal points allow three possible paths. The three $p_k(t)$ paths and the corresponding three $q_j(t)$ paths have mixed paths $\boldsymbol{\alpha}=(\alpha_1, \alpha_2, \alpha_3)$, $\boldsymbol{\beta}=(\beta_1, \beta_2, \beta_3)$, and an action matrix $S[p_j, q_k] = S_{jk}$. $A$ and $B$ attempt to extremize the generalized action $\boldsymbol{\alpha}^T\underline{S}\boldsymbol{\beta}$ by selecting optimal distributions $\boldsymbol{\alpha}_0$ and $\boldsymbol{\beta}_0$ such that the generalized action is extremized.

I now show the generalization of Hamilton's principle that governs the form of the actual or "best" pair of mixed paths $\alpha[p]$ and $\beta[q]$. Because any path, indexed by $j$ for $A$ and $k$ for $B$, can be represented in terms of $p(t)$ or $q(t)$, there is the same number $n$ of possible $p$ paths and $q$ paths in a set of $n$ paths, as shown in Figure 2 for three paths. Corresponding to each $p_j(t)$ or $q_k(t)$ is $\alpha[p_j]=\alpha_j$ and $\beta[q_k]=\beta_k$. I represent these mixed paths with the vectors $\boldsymbol{\alpha}=(\alpha_1, \alpha_2, ..., \alpha_n)$ and $\boldsymbol{\beta}=(\beta_1, \beta_2, ..., \beta_n)$. I can also define a square matrix for the action $S_{jk}$ corresponding to all possible actions for the $n$ paths with given $q_i$ and $q_f$. The elements of this action matrix $S_{jk}$ are real values of the action evaluated at paths $p_j$ and $q_k$, i.e., $S[p_j, q_k] = S_{jk}$. In the three-path case, shown in Figure 2, the action matrix $\underline{S}$ is written

$$\underline{S} = \begin{Vmatrix} S_{11} & S_{12} & S_{13} \\ S_{21} & S_{22} & S_{23} \\ S_{31} & S_{32} & S_{33} \end{Vmatrix}.$$

I combine $\boldsymbol{\alpha}$, $\boldsymbol{\beta}$, and $S_{jk}$ to form the generalized action $\boldsymbol{\alpha}^T\underline{S}\boldsymbol{\beta}$ where $\boldsymbol{\alpha}^T$ is the transpose of $\boldsymbol{\alpha}$. In analogy to the case with perfect information, I extremize this generalized action by finding optimal mixed paths $\boldsymbol{\alpha}_0$ and $\boldsymbol{\beta}_0$. $A$ selects $\boldsymbol{\alpha}_0$ so as to extremize $\boldsymbol{\alpha}_0^T\underline{S}\boldsymbol{\beta}$ for any given $\boldsymbol{\beta}$, and $B$ selects $\boldsymbol{\beta}_0$ to extremize $\boldsymbol{\alpha}^T\underline{S}\boldsymbol{\beta}_0$ for any given $\boldsymbol{\alpha}$. $\boldsymbol{\alpha}_0$ and $\boldsymbol{\beta}_0$ are the actual distributions. If $p_i$ and $p_f$ are given, I can similarly define both $\alpha[p]$ and $\beta[q]$, and the two optimal distributions $\boldsymbol{\alpha}_0$ and $\boldsymbol{\beta}_0$ will extremize $\boldsymbol{\alpha}^T\underline{R}\boldsymbol{\beta}$.

As shown in ref [6] there is a stationary result for a system with $n$ possible paths. This extrema has $\boldsymbol{\alpha}_0$ parallel to $\underline{S}\boldsymbol{\beta}_0$, all components of $\boldsymbol{\alpha}_0$ are equal, and all components of $\boldsymbol{\beta}_0$ are equal. To get this solution, I assume that a given set of paths has the same probability to be observed independently of whether they are represented in terms of $p$ or $q$. This extrema has an analogy in the mini-max extrema of zero sum games [7]. Unlike a zero-sum game, however, the generalized form of Hamilton's principle may have solutions where individual elements of an optimal vector may be negative, positive, or anti-communitive. The element, e.g., $S_{12}$ may be large and $S_{12}\beta_2$ would seem to be a large potential positive payoff to $B$. $A$'s element $\alpha_1$, however, may be negative to create a potentially large loss to $B$ $\alpha_1 S_{12}\beta_2$. This concept of the mixed path allows $A$ or $B$ to select a negative value for an element of their mixed path in order to "negate" a large expected value of the action for the other experimenter. $A$ or $B$ may be attracted or repelled from a path because of the large expected value of the generalized action that the other "player" may get. The game analogy also opens the door for speculation concerning a conceptual relation to competitive behavior in complex systems, biology, economics, etc. It may also be possible to apply computational methods from this area.

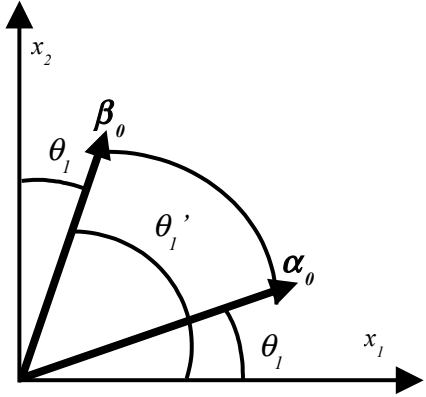

Figure 3. By combining the two optimal vectors $\alpha_0$ and $\beta_0$ in a common basis using two angles $\theta_1$ and $\theta_1'$, where $\theta_1 + \theta_1' = \pi/2$ are the angles from $\alpha_0$ and $\beta_0$ to the $x_1$ axis and $\theta_2 + \theta_2' = \pi/2$ are the angles to the $x_2$ axis, I can then construct a complex quantity for path 1 and path 2, i.e., $\phi_1 = \alpha_1 + i\beta_1 = |a|\cos(\theta_1) + i|a|\sin(\theta_1) = |a|\exp(i\theta_1)$ and $\phi_2 = \alpha_2 + i\beta_2 = |a|\cos(\theta_2) + i|a|\sin(\theta_2) = |a|\exp(i\theta_2)$. $\phi_1$ and $\phi_2$ are the amplitudes for path *1* and *2* respectively.

Even though $\alpha_0$ is parallel to $\underline{S}\beta_0$, $\alpha_0$ is parallel to $\beta_0$ only in the extraordinary case when $\underline{S}$ is diagonal. I may combine these two vectors using common parameters if they are expressed in a common basis. In the 2-D case, as shown in Figure 3, $\{\alpha_1, \beta_1\}$ is $\{|a|\cos(\theta_1), |a|\sin(\theta_1)\}$ and $\{\alpha_2, \beta_2\}$ is $\{|a|\cos(\theta_2), |a|\sin(\theta_2)\}$, where $\theta_1 + \theta_1' = \pi/2$ and $\theta_2 + \theta_2' = \pi/2$. The angles $\theta_2$ and $\theta_2'$ specify the angle between $\alpha_0$ and $\beta_0$ and the $x_2$-axis. This same analysis is easily generalized to many paths, e.g., for three paths the two vectors with a given angle between them can be oriented with 3 angles in a 3 dimensional vector space that is fixed by three constraints between the corresponding angles (or direction cosines), $\theta_i + \theta_i' = \pi/2, i=1,2,3$. In $n$ dimensions there are $n$ paths and $n$ constraints between corresponding angles. I can then find the total probability amplitude $K(q_i,t_i;q_f,t_f)$ as the sum over all $\phi_j$, i.e., $K(q_i,t_i;q_f,t_f) = \Sigma\phi_j = a\Sigma\exp(i\theta_j)$ between the given endpoints. In the quantum mechanical domain $\theta_j=(2\pi/h)S_j$, where $h$ is Planck's constant, and the propagator is $K(q_i,t_i;q_f,t_f)$ [8]. The usual argument for the additivity of $S_j$ between consecutive endpoints implies the multiplying of $K$'s between consecutive endpoints. This in turn leads to the Schrödinger Wave equation and is the essential ingredient for the path integral formulation of quantum mechanics [8] and quantum field theory [9]. In this case, the Heisenberg uncertainty principle $h \leq \delta p \delta q$ gives us the measure of the uncertainty for phase space points and tells us the lower limit of imperfect information, see the discussion in reference [6]. This formulation, however, is more general and should also be applicable in cases when $\delta p \delta q >> h$.

A new connection between classical mechanics and quantum mechanics is proposed. The quantum mechanical propagator is derived from a generalized form of Hamilton's principle. The system has imperfect information and two distribution functionals over possible $p$ paths $\alpha[p(t)]$, over possible $q$ paths $\beta[q(t)]$, and a generalized action corresponding to a matrix of the action evaluated at all possible $p$ and $q$ are defined. The generalized Hamilton's principle is the extremization over all possible distributions of $\iint_{p,q} DqDp\alpha[p]\,S[p,q]\,\beta[q]$. The normalization of the distributions allows their values to be real numbers between $+1$ and $–1$ for bosons and Grassman numbers (between $+1$ and $–1$) for fermions. The two optimal distributions are identified as the real and imaginary parts of the complex amplitude.